\documentclass{article}
\usepackage[left=1in, right=1in]{geometry}
\usepackage[utf8]{inputenc}
\usepackage{graphicx}
\usepackage{tikz}

\newcommand*\annotatedFigureText[4]{\node[draw=none, anchor=south west, text=#2, inner sep=0, text width=#3\linewidth,font=\sffamily] at (#1){#4};}
\newenvironment {annotatedFigure}[1]{\centering\begin{tikzpicture}
\node[anchor=south west,inner sep=0] (image) at (0,0) { #1};\begin{scope}[x={(image.south east)},y={(image.north west)}]}{\end{scope}\end{tikzpicture}}
\usepackage{fancyhdr}
\usepackage{wrapfig}
\usepackage{gensymb}
\usepackage{mathtools}
\usepackage{ulem}
\graphicspath{{images/}}
\usepackage{amsmath}
\usepackage{bbm}
\usepackage{mathrsfs}
\usepackage{cancel}
\usepackage{amsfonts, amsthm}
\title{\vspace{-1.0in} Characterizing Turbulence at a Forest Edge: A Vorticity Budget Analysis around a Canopy}
\author{Dorianis M. Perez$^{123}$, Jesse M. Canfield $^{3}$, Rodman R. Linn $^{4}$, Kevin Speer $^{12}$}
\date{}

\begin{document}
\maketitle
\vspace{-0.3in}
\begin{center}
\hspace{-0.1in} \textit{$^{1}$Geophysical Fluid Dynamics Institute (GFDI), Florida State University, Tallahassee, FL} \\
\textit{$^{2}$Department of Scientific Computing (DSC), Florida State University, Tallahassee, FL} \\
\textit{$^{3}$X Computational Physics Division, Los Alamos National Laboratory, Los Alamos, NM}
\\
\textit{$^{4}$ Earth and Environmental Science Division, Los Alamos National Laboratory, Los Alamos, NM}
\end{center}

Corresponding author: Dorianis M. Perez dorianisp@lanl.gov

LA-UR-24-22033

\section*{Abstract}
Vorticity is a key characteristic of flow patterns that determine wildland fire behavior, frontal evolution, and wind-canopy interaction. Investigating the role of vorticity in the flow fields around vegetation can help us better understand fire-atmosphere feedback and the influences of vegetation on this feedback. In modeling vorticity, ``perhaps the greatest knowledge gap exists in understanding which terms in the vorticity equation dominate [...] (and) when one or the other might dominate" (Potter, 2012). In this study, we investigate the role of vorticity in boundary layer dynamics and canopy/forest edge effects using HIGRAD/FIRETEC, a three-dimensional, two-phase transport model that conserves mass, momentum, energy, and chemical species. A vorticity transport equation was derived and discretized. Simulations were performed over a cuboidal homogeneous canopy surrounded by surface vegetation. This derivation led to the discovery of a drag tilting and stretching term, which shows that gradients in the aerodynamic drag of the vegetation, tied to heterogeneities in surface area-to-volume ratio, play an important role in the generation of vorticity. Results from the vorticity budget analysis show that this term contributes significantly in the areas where these gradients are present, namely the edges of the canopy. 

\section*{Keywords}
Turbulence, Vorticity, Boundary Layer, Canopy

\section*{Introduction}
Vorticity is generated by wind shear, which is often tied to gradients in pressure, temperature,density, or aerodynamic resistances or blockages. If wind is blowing in any outdoor environment, there will be wind shear and thus vorticity produced near the ground due to surface drag, vegetation, and terrain features that provide obstacles in the flow (Forthofer and Goodrick 2011). In and around a canopy, an abundance of ambient vertical and horizontal vorticity is generated by strong wind shear associated with gradients in aerodynamic drag and especially the interaction between slowed flows within the canopy and fast moving air above the canopy (Sharples et al. 2022). Further understanding how vortices play a role in forest edge effects on flow patterns can help provide answers as to how the presence of a canopy can contribute to vortex generation and the implications this could have on a variety of phenomena including fire behavior. 

Wildfire behavior is directly related to wind flow dynamics, turbulence, and vorticity phenomena within and around the canopy (Pimont et al. 2009). Large coherent structures are known to dominate to turbulent transport of momentum around canopies (Dupont and Brunet 2009). Therefore, the effect of these developing coherent structures has implications on phenomena in forest canopies, namely wildland fire behavior. 
The two most commonly-noted vortex forms observed in wildland fire are fire whirls (spin about a vertical axis) and horizontal roll vortices (spin about a horizontal axis) (Forthofer and Goodrick, 2011). Fires interact with and modify ambient vorticity, but also generate additional vorticity. With a fire's pyroconvective updraft, vorticity can then be reoriented, often referred to as tilting and stretched (Cunningham et al. 2005; Forthofer and Goodrick 2011; Sharples et al. 2015; Vallis 2017; Tohidi et al. 2018).

When considering the dynamics that can alter the mean flow, there are numerous mechanisms that play a role. The vertical structure of the mean velocity field near and above the surface layer, the lowest part of the atmospheric boundary layer where surface drag-induced shear plays an important role, influences the dynamics of eddy generation, which affects the balance of mass, momentum, and heat surrounding the canopy (Bebieva et al. 2021; Pimont et al. 2022). In the inertial sublayer, above the ground and the canopy, it has been established that the combined effects of drag, shear turbulence production, and turbulence dissipation result in a typical logarithmic vertical velocity profile 
(Pimont et al. 2022). Below the inertial sublayer, the overall flow pattern is controlled by the presence of vegetation and depends on vegetation type, density, horizontal heterogeneity, and vertical profile. In most mesoscale weather models, the combined effects of these quantities is often parameterized as surface roughness (Bebieva et al. 2021; Kaimal and Finnigan 1994; Raupach et al. 1996), but these tools do not attempt to capture the flow within the vegetation or canopy layer. The canopy structure imposes a drag force on the wind field, which acts as a momentum sink. Canopy drag results in gradients and an inflection point in the vertical velocity profile near the top of the canopy and this inflection point leads to the development of Kelvin-Helmholtz instabilities, which roll over to form transverse vortices that are then transformed through secondary instabilities into three-dimensional structures (Dupont and Brunet, 2009, Holton and Hakim 2012). The development of coherent turbulent structures, or eddies, characterize the wind flow in and around the canopy (Bebieva et al. 2021). These eddies lead to downdrafts (sweeps) into the canopy and weaker updrafts (ejections) vertically out of the canopy, driving mixing between the slow moving winds within the canopy and the fast moving winds aloft. This sweep and ejection-driven mixing is responsible for most of the vertical momentum transfer between the canopy and atmosphere (Dupont and Brunet 2009; Gao et al. 1989; Lu and Fitzjarrald, 1994). Thus, the canopy plays a critical role in determining the magnitude and structure of wind flow near the surface, and consequently, on phenomena including fire behavior, moisture exchange, seed dispersal, and even energy balances of plant material. 

Coherent turbulent structures have been studied numerically for decades using both large eddy simulations (LES) and direct numerical simulations (DNS) (Kanda and Hino, 1994; Su et al. 2000; Watanabe, 2004; Shaw et al., 2006; Dupont and Brunet, 2008b; Finnigan et al., 2009; Dupont and Brunet, 2009). Compared to other computational fluid dynamics (CFD) models, LES has proven capable of resolving wind gusts in a plant canopy as well as large coherent eddy structures that may be crucial to wildfire propagation and has been successfully applied over various homogeneous and heterogeneous vegetation canopies under neutral stratification (Pimont et al. 2009; Shaw and Schumann 1992; Kanda and Hino 1994; Su et al., 1998, 2000; Watanabe 2004; Dupont and Brunet 2007, 2008b; Patton et al. 1998; Yang et al. 2006a, 2006b). Finnigan et al. (2009) presents one of the most comprehensive studies of coherent structures over homogeneous canopies, characterizing the structure of coherent eddies from a composite average deduced from LES outputs (Dupont and Brunet, 2009). 

One of the most important metrics in LES simulations of flow over canopies is the turbulent kinetic energy (TKE). Both large coherent structures and small-scale structures contribute to the total TKE in and around a canopy. TKE budgets highlight the various physical processes that dictate turbulent fluid motion in this region (Dwyer et al. 1997). One of the earliest studies of the canopy TKE budget was performed by Lesnik (1974), whose initial findings are still being supported by recent studies; mainly, that shear production is a major source at the canopy top where wind shear is the strongest and buoyancy is small in comparison to other terms (Dwyer et al. 1997). TKE budgets have been measured in a variety of canopies in the field, wind tunnels and numerically using LES (Leclerc et al. 1990; Dwyer et al. 1997; Meyers and Baldocchi 1991; Shaw and Seguner 1985; Raupach et al 1986; Brunet et al 1994; Finnigan 2000). However, vorticity budgets have yet to be investigated and characterized term-by-term. In association with wildland fire, vortices are often discussed in the context of plume dynamics, namely large fire whirls and horizontal roll vortices as in Forthofer and Goodrick (2011) and Potter (2012). Background vorticity from the wind field dynamics interacting with the vegetation has not been extensively described in the literature. 


The subject of the present paper is to investigate the role of vorticity in wind flow dynamics within and around an idealized forest canopy using HIGRAD/FIRETEC. We show results of a quantitative spatial- and time-averaged vorticity budget. Specifically, we investigate vorticity dynamics at the interface between a canopy and open grassland. A detailed validation study of wind field/canopy interactions in HIGRAD/FIRETEC shows excellent agreement with experiments (Pimont et al., 2009). Potter (2012) recalls that even the most detailed study of vortex generation by wildfire stops short of using the numerical model data to examine the various terms in the vorticity equation quantitatively. In order to ease the derivation of the vorticity budget equation outlined in this study, we used a version of HIGRAD/FIRETEC that included the Smagorinsky eddy viscosity model. The performance of the Smagorinsky model in the HIGRAD/FIRETEC framework was proven to provide reliable physics when compared to the original 1.5-order closure model posed in HIGRAD/FIRETEC in \textit{Perez et al. (2024) (submitted)}. Thus, we proceed with the consideration of the Smagorinsky residual stress tensor in the momentum equation (Equation \ref{momentum}). 


\subsection*{Model Description}
HIGRAD/FIRETEC is a three-dimensional coupled fire-atmosphere model that conserves mass, momentum, energy, and chemical species written in terrain-following coordinates (Pimont et al. 2009). The equations and corresponding terms are defined in detail in Linn (1997), Linn et al. (2002), and Linn and Cunningham (2005). FIRETEC is composed of physics models that predict atmosphere fuel interactions, combustion, heat transfer via convection and radiation, emission of particles like black and organic carbon, and generation and transport of firebrands (Linn et al. 2012; Colman and Linn, 2007; Linn et al. 2005; Josephson et al., 2020; Koo et al. 2012). HIGRAD computes compressible fluid flow in the lower atmospheres, solving Navier-Stokes equations with an LES approach (Reisner et al. 2000a, 2000b). HIGRAD solves the Navier-Stokes equations using a forward-in-time numerical technique and details can be found in Reisner et al. (2000a, 2000b). The momentum equation in the version of HIGRAD/FIRETEC with the Smagorinsky model implemented is designed to separate small and large scales and is written as

\begin{equation}
    \rho \Big( \frac{\partial \textbf{u}}{\partial t}  + \textbf{u} \cdot \nabla \textbf{u} \Big)  = - \nabla p + \rho \textbf{g} - \nabla \cdot \boldsymbol{\tau} - \rho C_D a_v |u| \textbf{u}
    \label{momentum}
\end{equation}

\noindent where $\rho$ is the combined gas density, $\textbf{u}$ is the combined gas velocity, $ \textbf{g}$ is the gravitational acceleration, $\boldsymbol{\tau}$ is the residual stress tensor, $C_D$ is the vegetation drag coefficient. Turbulent structures larger than the largest specified turbulent length scale are explicitly solved by the model, and subfilter-scale (SFS) turbulent structures are modeled. In the context of wildfires, complex SFS turbulent kinetic energy (TKE) transformations occur at various spatial and temporal scales within subfilters. The turbulent residual stress tensor, $\boldsymbol{\tau}$, is modeled using the Smagorinsky eddy viscosity model, $\boldsymbol{\tau} = -2 \rho \nu \textbf{S}$, where $\nu = \ell^{2} \overline{S}$ is the modelled eddy viscosity. $\overline{S}$ is the characteristic filtered rate of strain and $\ell = C_{S}\delta$ represents the Smagorinsky length scale which is proportional to the filter width, $\delta = \sqrt{\Delta x^2 + \Delta y^2 + \Delta z^2}$, through the Smagorinsky coefficient, $C_{S} = 0.17$ (Pope, 2000). $\Delta x$, $\Delta y$, and $\Delta z$ are the grid resolution in the $x,y,z$ directions, respectively.

We model a forest canopy in HIGRAD/FIRETEC as a porous media that imparts drag on the momentum field in computational cells that contain the canopy. 
The momentum drag term has an engineering drag coefficient, $C_D$, the surface area to volume ratio of fuel elements in a computational volume, $a_v$, and the flow velocity, $u$. The vegetation density (fuel density) is a function of the variable $a_v$. We take the curl of the momentum equation to yield the vorticity transport equation presented in the next section.

\subsubsection*{Vorticity Transport Equation}
By definition, vorticity is the curl of the flow velocity and is expressed as $\vec{\omega} = \nabla \times \vec{u}$ (Pope, 2000). By taking the curl of the momentum equation as it is expressed in Equation \ref{momentum}, we derive the vorticity transport equation of interest

\begin{equation}
  \begin{split}
    \frac{\partial \boldsymbol{\omega}}{\partial t} + \underbrace{(\mathbf{u} \cdot \nabla) \boldsymbol{\omega}}_{\omega_\text{adv}}  =  \underbrace{( \boldsymbol{\omega} \cdot \nabla) \mathbf{u}}_{\omega_\text{vts}} - \underbrace{\boldsymbol{\omega}(\nabla \cdot \mathbf{u})}_{\omega_\text{div}}  \\ 
     & \hspace{-1.5in}- \underbrace{\frac{1}{\rho^2} \nabla \rho \times \nabla p}_{\omega_\text{bar}} -  \underbrace{ \frac{1}{\rho^2} \nabla \rho \times \nabla \boldsymbol{\tau}}_{\omega_\text{visc}} - \underbrace{C_D (\nabla  a_v|\mathbf{u}|) \times \mathbf{u}}_{\omega_\text{dts}} -  \underbrace{C_D  a_v |\mathbf{u}|  \boldsymbol{\omega }}_{\omega_\text{dshear}} 
  \end{split}
  \label{VTE}
\end{equation}

\noindent where second term on the left side, $(\mathbf{u} \cdot \nabla) \boldsymbol{\omega}$, is the advection of vorticity by resolved flow, the first term on the right side, $( \boldsymbol{\omega} \cdot \nabla) \mathbf{u}$, represents vortex realignment, tilting, and stretching and describes how velocity gradients can transform horizontal vorticity into vertical vorticity and vice versa (Forthofer and Goodrick 2011). The second term,  $\boldsymbol{\omega}(\nabla \cdot \mathbf{u})$, represents flow convergence/divergence and how vortices are stretched/compressed and increases/decreases the magnitude of the vorticity (Forthofer and Goodrick 2011). The third term is baroclinicity, $\frac{1}{\rho^2} \nabla \rho \times \nabla p$, which generates vorticity in cases where pressure and density gradients are not parallel (Forthofer and Goodrick 2011). In wildfires, heating at the surface causes a horizontal temperature gradient that is not aligned with the ambient vertical pressure gradient. This misalignment is what causes rotational motions to mix warm and cold fluid in an attempt to restore balance (Forthofer and Goodrick 2011). The fourth term, $\frac{1}{\rho^2} \nabla \rho \times \nabla \boldsymbol{\tau}$, represents generation of vorticity from viscous shear stress (Forthofer and Goodrick 2011). Horizontal vorticity is being generated by vegetation of ground surface drag-induced wind shear (Forthofer and Goodrick 2011). The gravitational acceleration is not considered here. The viscous shear stress here is the residual stress tensor-eddy viscosity relation represented through the Smagorinsky model described and analyzed in \textit{Perez et al. (2024), submitted}.

The last two terms in Equation \ref{VTE} originate from the momentum drag term. The first drag term, $C_D (\nabla  a_v|\mathbf{u}|) \times \mathbf{u}$ , describes the change in direction of vorticity, $|\textbf{u}|$, due to velocity interactions with gradients in the drag imposed by vegetation bulk density, which is proportional to the bulk surface area-to-volume ratio of the vegetation, $a_v$. This drag can could be expressed in other similar variables such as the Leaf Area Index (LAI), which are also proportional to local bulk density of the vegetation. In this term, $|\mathbf{u}|$ scales $a_v$ so that, to leading order, the vorticity increases quadratically with the local wind velocity, $\boldsymbol{\omega} \approx -C_D (\nabla  a_v) \times \mathbf{u} |\mathbf{u}|$.  This term serves predominantly as a vorticity source because the gradient in bulk vegetation density (and thus surface area per unit volume) tends to coincide with a velocity gradient of opposite sign. We derive this term from the momentum equation that includes vegetation drag terms and refer to it as the drag tilting and stretching term. This term has the physical effect of tilting vortex sheets into the direction that is normal to the gradient of $a_v$ and the direction of the velocity field. It also exhibits vortex stretching where the vorticity concentrates in regions where there is a gradient in $a_v$. The second of the two drag terms in Equation \ref{VTE} is a vorticity sink that develops in locations of vegetation bulk density gradients. This sink is largest at canopy edges. These boundary layers produce shear in the velocity field and manifest as vortex sheets normal to the velocity gradient.

The vorticity transport equation is discretized and solved term-by-term from the computed resolved variables in HIGRAD/FIRETEC, uncluding velocity, surface area per unit volume, gas density, and now vorticity. We present a preliminary overall budget of the vorticity transport equation through time with streamwise wind flow through a canopy, as well as a deeper look into the drag terms to study their role at the edges of the homogeneous canopy. We will refer to the vector components of velocity and vorticity with the following nomenclature, $\textbf{u} = (u, v, w)$, and $\boldsymbol{\omega} = (\xi, \eta, \zeta)$, for the x-, y-, and z-components respectively. We present numerical details of setting up simulations to test this term and further analyze its results in the following sections.

\section*{Numerical Details}
\begin{figure}[h!]
         \centering
         \includegraphics[width=10cm]{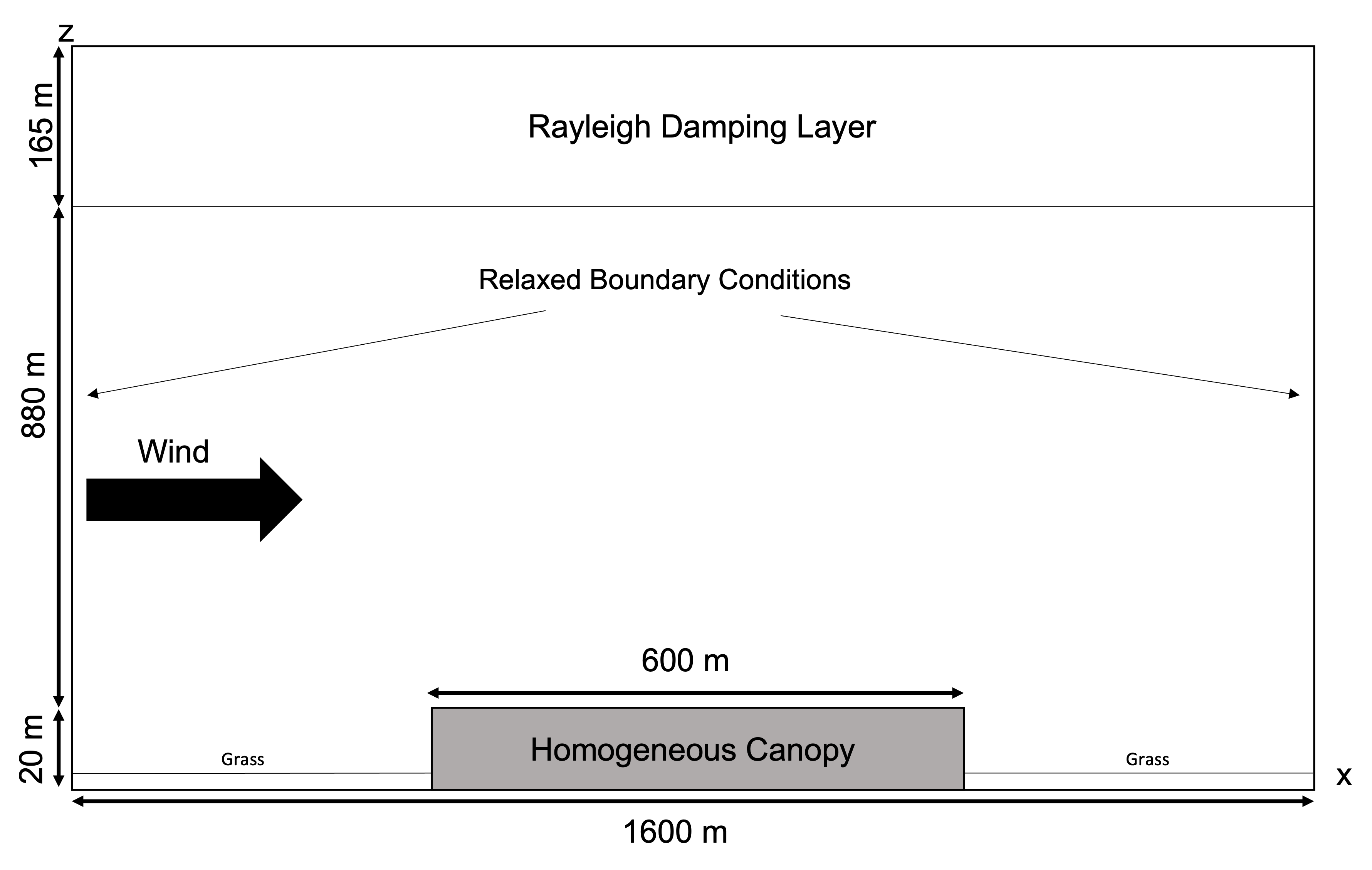}
         \caption{Schematic representation of the computational domain used in homogeneous canopy simulations}
         \label{schematic}
\end{figure}

Pimont et al. (2009) validated the wind dynamics of HIGRAD/FIRETEC over vegetated canopies with a neutral thermal stratification and without considering fire propagation or interaction between winds and strong buoyant sources. The model was validated over a continuous forest canopy using the in-situ measurements of Shaw et al (1988) and a discontinuous forest using the wind-tunnel measurements of Raupach et al. (1987) (Pimont et al. 2009). With findings and recommendations from Pimont et al. (2006; 2009) and Canfield et al. (2005) considered in the present study, the computational domain size is 1600 X 1600 X 1065m with a horizontal spacing of $\Delta x = \Delta y =2m$ and vertical mesh stretching, $\Delta z$ starting from 1.5m near the ground to 40m at the top (Figure \ref{schematic}). The same domain was utilized in \textit{Perez et al. (2024), submitted}.

The canopy height was set at 20m. The chosen domain size was selected to be large enough to include the near surface eddies induced by local surface/vegetation interactions (in the context of an idealized infinitely long flat surface expanse) where the size of the largest eddies of this sort is determined by defining an integral length scale that is a measure of the largest correlation distance between two points in the turbulent flow. Pre-computed wind fields with cyclic boundary conditions in the x- and y-directions were used as inflow boundary conditions on the x inlet, as suggested by Canfield et al. (2005) and Pimont et al. (2009). A Rayleigh damping layer was used at the upper boundary and in the vertical direction, free-slip, non-penetration condition is on the lower boundary (Pimont et al. 2009). The size of the computational domain was chosen so that the boundaries have a negligible effect on the dynamics of the flow around the canopy and sensitivity tests were performed to ensure that the boundaries are placed sufficiently far.

\section*{Results and Discussion}
\subsubsection*{The Drag Tilting and Stretching Vorticity}
The derivation of Equation \ref{VTE} revealed two vorticity drag terms. The drag tilting and stretching term, $\omega_\text{dts}$,  describes the change in magnitude and direction of the vorticity vector due to the cross product of the drag amplitude gradient and the velocity vector, 
\begin{equation}
    \omega_\text{dts} =  C_D \Bigg( \bigg(\frac{\partial a_v |u|}{\partial y} w - \frac{\partial a_v |u|}{\partial z}v \bigg) \mathbf{i} + \bigg( \frac{\partial a_v |u|}{\partial z}u - \frac{\partial a_v |u|}{\partial x} w \bigg) \mathbf{j} + \bigg(\frac{\partial a_v |u|}{\partial x} v - \frac{\partial a_v |u|}{\partial y}u \bigg) \mathbf{k} \Bigg)
    \label{dragtilt}
\end{equation}

\noindent The resulting vorticity is orthogonal to the vegetation gradient and the velocity vector. Thus, the term is expected to be large at the edges of our idealized canopy, where the vegetation gradient is sharp. 

\begin{figure}[h!]
        \centering
        \includegraphics[width=0.65\textwidth]{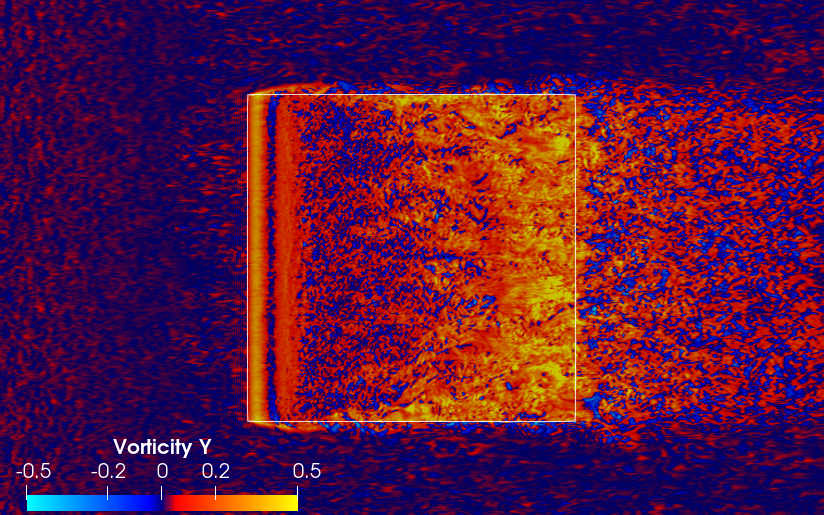}
        \includegraphics[width=0.65\textwidth]{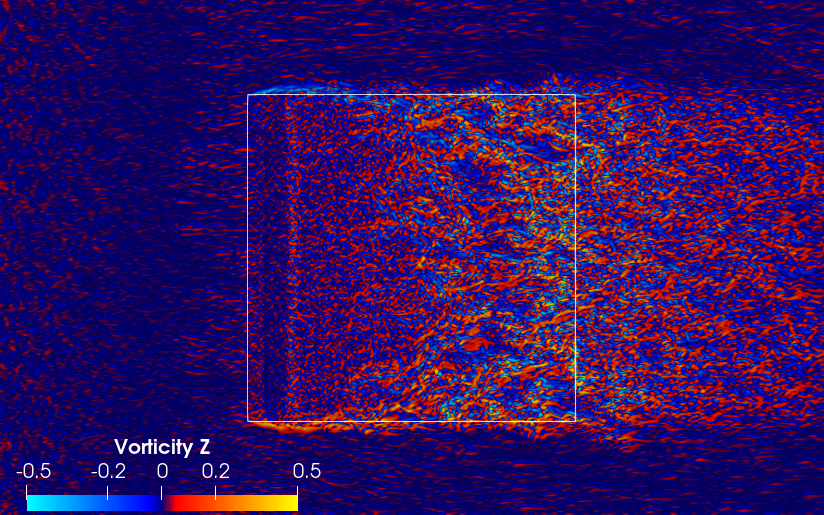}
         \caption{Horizontal cross-sections contoured by instantaneous $\eta$ (top), and $\zeta$ (bottom), of the vorticity along the top surface of the canopy at time 2500 seconds.}
         \label{isosurface_vort}
\end{figure}

Figure \ref{isosurface_vort} shows color contours of vorticity components in an x-y plane at the top surface of the canopy where the the mean flow is from left to right (x-direction) and the canopy region is outlined with the white line. The y-component (cross-stream component), $\eta$, (top) and z-component (vertical component), $\zeta$ (bottom) of the vorticity vector are colored by their magnitudes. A high concentration of $\eta$ is observed at the leading edge of the canopy (left side of the canopy square) that is generated by the drag tilting and stretching term. In the x-direction, there is a positive $a_v$ x-direction gradient at the leading edge and a negative vertical $a_v$ gradient at the trailing edge of the canopy. The wind field at the leading edge has a vertical component as the air is flowing up and over the canopy. Thus, the vertical velocity, $w$, is positive and the stream-wise velocity, $u$, is also positive. The two products of the $a_v$ gradients and the velocity components work together to provide a source term for positive $\eta$. The large area of high $\eta$ that develops about a quarter of the distance downstream of the leading edge represents vortex development from the shear layer over the canopy. The main contribution of the vorticity in these areas is the product of the vertical gradient of $a_v$ and $u$. In the $\zeta$ plot (Figure \ref{isosurface_vort}, bottom) the highest concentrations of vorticity are on the two lateral flanks of the canopy, where there is a sharp y-gradient in $a_v$, normal to the velocity. The top lateral edge (positive y-direction side of left side if you are looking in the direction of the mean flow) shows negative $\zeta$ and the bottom lateral edge shows positive $\zeta$. The lateral edges with higher magnitude $\zeta$ support that vorticity is tilted into the vertical direction where there is a sharp y-gradient in $a_v$. 

\begin{figure}[h!]
        \centering
        \includegraphics[width=0.65\textwidth]{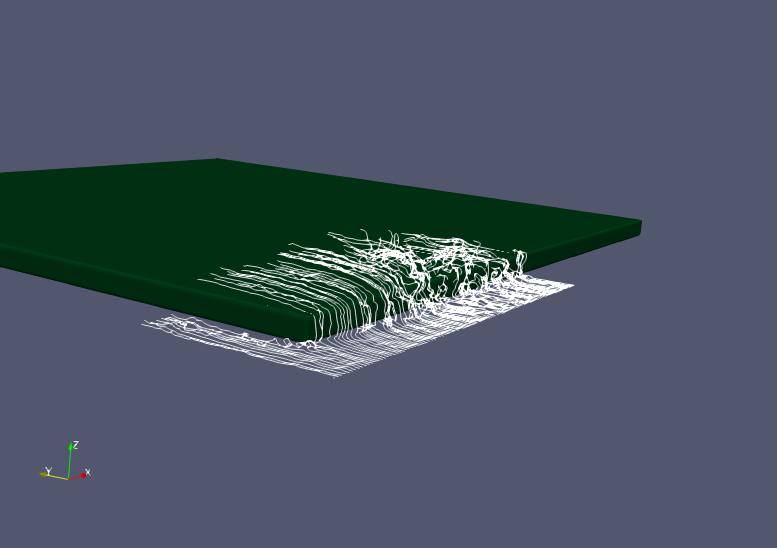}
        \includegraphics[width=0.65\textwidth]{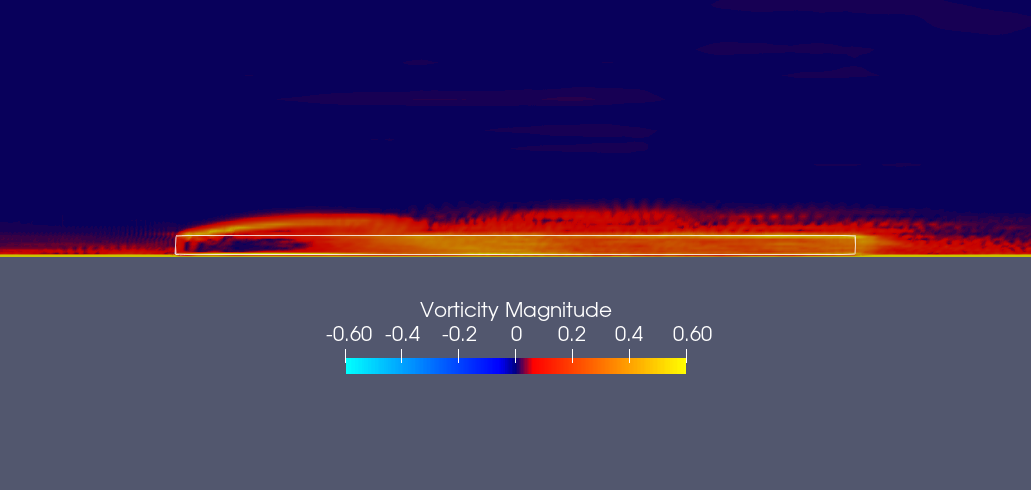}
         \caption{Vortex lines parallel to the spanwise direction at the lateral edge of the canopy showing tilting and stretching induced by the drag (top). This tilting of vorticity at the lateral edge is due to the sharp gradient in $a_v$. The sharp gradient $a_{v,y}$ at the lateral edge and $a_{v,x}$ at the leading edge contribute large amounts of vorticity, as can be visualized in the vertical cross-section of vorticity magnitude adjacent to the lateral edge of the canopy (bottom).}
         \label{streamline_vort}
\end{figure}

The orientation of the vorticity with respect to the mean flow (largely in the x direction) and the shear associated with the edge of the canopy can be observed in Figure \ref{streamline_vort} (top) with streamlines to visualize the direction of the vorticity vector. That is, the streamlines are aligned with the axis of rotation for vortical motion. Hence, we will refer to these streamlines as vortex lines. In order to highlight the alignment of the vorticity and influence of vegetation, the white box encloses a small volume that contains the leading, top, and one of the lateral edges of the canopy, shown as a green isosurface. The vortex lines are contained in the white box so that they don't saturate the rest of the graphic. The axis in the lower left of the image shows the Cartesian directions. Upstream of the canopy, the vortex lines are approximately straight in the y-direction, indicating an $\eta$ vortex sheet at the top surface of the grass. The shear profile in the mean flow drives these $\eta$-vortex sheets. The sink of momentum by drag dissipates energy in the layer of grass, decreasing the wind field at the bottom. This effect results in a larger vertical velocity gradient that concentrates the vorticity at the top of the grass, where the vegetation gradient is sharp. At the corner and along the lateral edge of the canopy, the vortex lines are realigned or tilted into the vertical direction. Furthermore, the vortex lines conform to the canopy profile. As they are deformed around the canopy perimeter along with the mean flow, they are elongated or stretched. Over the top of the canopy the vortex lines are again realigned or tilted back into the y-direction along the top surface (perpendicular to the main velocity gradient). Farther downstream the vortex lines are more chaotic because they have been tilted, lifted, and stretched with the development of the turbulent boundary layer that develops above and around the canopy. 

Color contours of vorticity magnitude (Figure \ref{streamline_vort}, bottom) along the lateral edge shows the highest values streaming off of the leading edge as the mean flow tilting and stretching velocity gradients are combined with the drag tilting and stretching contributions to $\eta$ and are then advected with the flow, $\omega_{vts}$. On the lateral edge high values of vorticity are due to $\zeta$, induced by the same term. To the authors' knowledge, this result has yet to be described in the literature. These results have implications for the enhancement of fire whirls along gradients in the fuel density that will be further described in the Conclusion section.

\begin{figure}[h!]
        \centering
        \includegraphics[width=11cm]{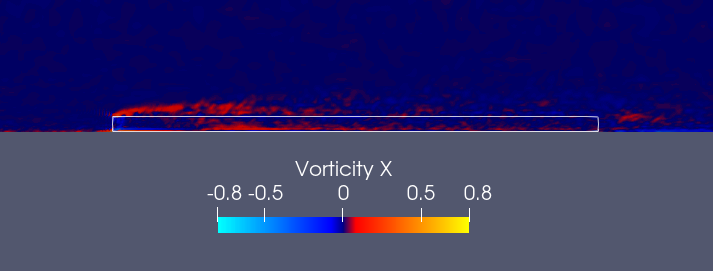}
        \includegraphics[width=11cm]{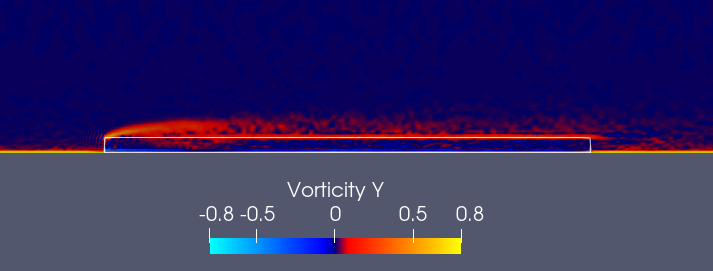}
        \includegraphics[width=11cm]{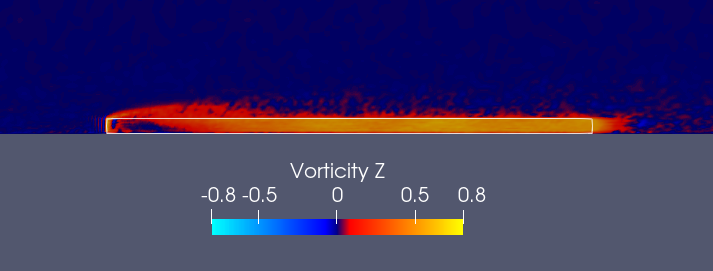}
        \caption{Vertical cross-sections of the time-averaged x-component of vorticity, $\xi$,  y-component of the vorticity, $\eta$, and z-component of the vorticity, $\zeta$, adjacent to the lateral edge. $\zeta$ shows a higher magnitude due to the combined effects of the gradient of $a_v$ in the x- and y-direction.}
        \label{vortcomponents}
\end{figure}

Color contours of the time-averaged vorticity components are depicted in Figure \ref{vortcomponents}. Upwind of the canopy, the vorticity components are relatively similar in magnitude, except for the y-component, $\eta$, which presents higher magnitudes at the top of the grass layer as described above. At the leading edge, $\eta$ is much higher in magnitude, reaching a maximum at the top of the leading edge in the area of positive vertical velocity. This is driven by the drag tilting and shear terms as described in Figure \ref{isosurface_vort}. Above the canopy, $\eta$ is consistently larger in magnitude than the x-component, $\xi$, because of the contribution from the mean wind shear, which forms in the shear layer of above the canopy above the canopy that develops and propagates downwind of the leading edge through to the trailing edge of the canopy. This is consistent with results from Dupont and Brunet (2009). The z-component, $\zeta$, has the highest magnitude of all components at the lateral edge of the canopy. We expect this result, as this is precisely where the greatest velocity gradients are oriented in the y-direction and vorticity is oriented vertically at the corner of the canopy. The sharp gradient of $\frac{\partial a_v}{\partial x}$ at the leading corners of the canopy contribute to the generation of vorticity that is advected along the lateral edge and the sharp y gradient in velocity and gradient of $\frac{\partial a_v}{\partial y}$ further contribute to the generation  vertically-oriented vorticity along these lateral sides. 

\subsubsection*{Enstrophy}
Enstrophy is a scalar quantity that may indicate potential energy contained in vorticity. It can also serve as a proxy for the magnitude of the vorticity because it is a mean square of the vorticity components. However, areas of high enstrophy do not necessarily imply the presence of a three-dimensional vortical structurse because it is a scalar. The enstrophy is a measure of the local rotation rate of the flow, defined as the vorticity squared, $\boldsymbol{\omega} \cdot \boldsymbol{\omega}$,
and represents a scalar quantity that intrinsically reflects the strength of the vorticity field without a vector implication (Pope 2000; Denaro, 2018). It is calculated as half the sum of the squares of the three vorticity components, $\xi$, $\eta$, and $\zeta$, respectively (Dupont and Brunet 2009). Enstrophy is analogous to kinetic energy and the momentum vector in that the vorticity squared represents a scalar quantity that serves as a measure for the vortical energy of the turbulent structures (Pope 2000; Denaro 2018; Saeedipour 2023). Weiss (1990) found that the growth of spatial gradients in the vorticity scalar is related to the transfer of enstrophy to the Kolmolgorov-scale motion and that increased dissipation is caused by the transfer of enstrophy to the these scales. This is echoed in various numerical and laboratory experiments  concluding that high enstrophy is associated with a high dissipation rate of energy associated with rotational flows (Zhu and Antonia,  1997). 



\begin{figure}[h!]
        \centering
        \includegraphics[width=0.75\textwidth]{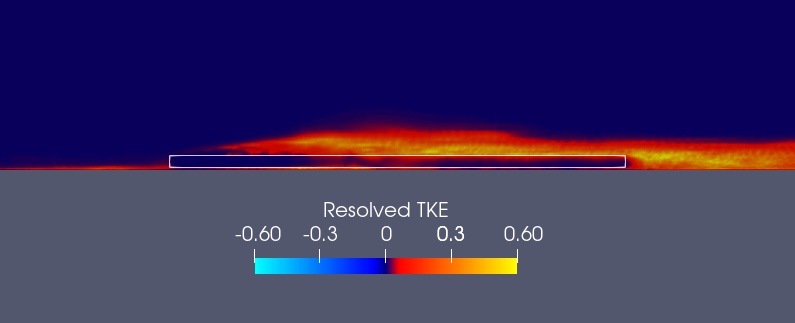}
        \includegraphics[width=0.75\textwidth]{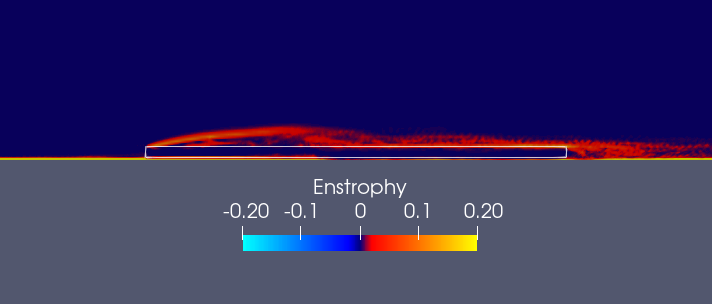}
         \caption{Time-averaged vertical cross-sections of the resolved turbulent kinetic energy (top) and the enstrophy (bottom) through the center line of the canopy. The resolved TKE emphasizes large scale turbulent motions and the enstrophy emphasizes small scale motions. Large scale motions are expected in the shear layer and downwind of the trailing edge.}
         \label{energy_enstrophycontour}
\end{figure}

\noindent Figure \ref{energy_enstrophycontour} shows a comparison of time-averaged x-z contours of the turbulent kinetic energy (top) and the enstrophy (bottom) of the resolved flow through the vertical x-z plane though the center of the canopy. This figure highlights the differences between the kinetic energy at the large resolved scales, and the enstrophy, an indicator of energy at the small resolved scales. Regions of high turbulent kinetic energy are expected to coincide with coherent structures. The adjustment region downwind of the Kelvin-Helmholtz waves show higher turbulent kinetic energy than enstrophy, and this region contains large coherent structures that will be discussed below. The shear boundary layer contains maximum values of resolved TKE, as the resolved-scale motions cause the vertical growth of the boundary layer downwind of the leading edge, above the canopy. Downwind of the trailing edge, is another area of high resolved TKE, where there are large vortices circulating adjacent to the canopy. The presence of a large recirculating vortex in this region is consistent with findings from flow over an obstacle in engineering applications (Behnamed and Aliane, 2019; Schroder et al., 2020). 

Smaller structures are expected in locations of high enstrophy. The enstrophy is high along the grass, leading up to the canopy and starts to increase near the top of the leading edge. The shear layer that develops in and above the grass contains small-scale vorticity that manifest as a layer of vortex sheets, confirmed by the maximum values of enstrophy in this location. At the top of the canopy, the enstrophy begins to increase and highlights the small structures in the turbulent boundary layer. These results are consistent with findings of areas of high enstrophy in and near a canopy from Dupont and Brunet (2009). 




\subsubsection*{Q-Criterion}

\begin{figure}[h!]
        \centering
        \includegraphics[width=10cm]{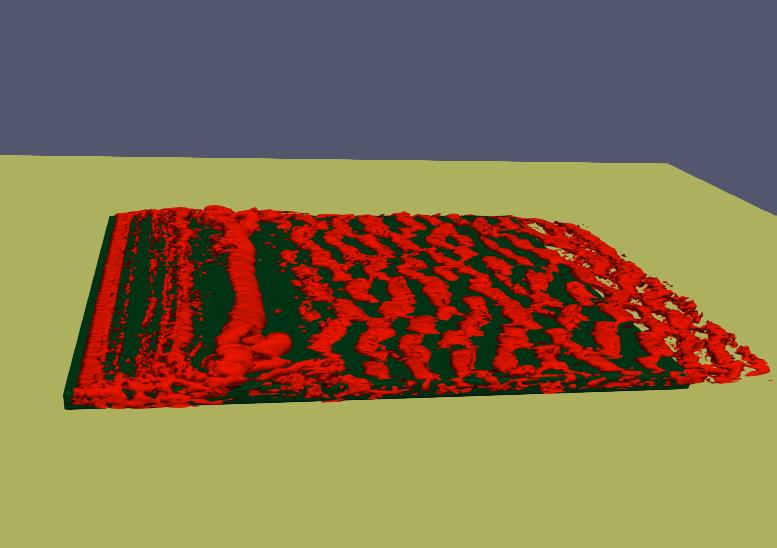}
         \caption{Isosurface of time-averaged positive values of the Q-criterion over the canopy showing three-dimensional coherent structures. Near the leading edge, we see a discernible pattern consistent with the formation of vortex tubes from the initial waves that are then broken up further downwind.}
         \label{qcriterion}
\end{figure}

The Q-Criterion quantifies the relative amplitude of the rotation rate and the strain rate of the flow and helps identify vortex cores. The Q-Criterion is expressed as,
\begin{equation}
    Q_c = \frac{1}{2} (\Omega_{ij} \Omega_{ij} - S_{ij} S_{ij}),
\end{equation}
where,
\begin{equation}
    \Omega_{ij} = \frac{1}{2}\bigg(\frac{\partial u_i}{\partial x_j} - \frac{\partial u_j}{\partial x_i} \bigg),
\end{equation}
and,
\begin{equation}
    S_{ij} = \frac{1}{2}\bigg(\frac{\partial u_i}{\partial x_j} + \frac{\partial u_j}{\partial x_i} \bigg),
\end{equation}
are the antisymmetric (rotation rate) and symmetric aspects (strain rate) of the velocity-gradient tensor, respectively. Thus, in regions where the rotation rate dominates the strain rate ($\Omega^2_{ij} > S^2_{ij}$),  the Q-criterion is positive and indicates the presence of vortex cores. Positive Q-criterion identifies regions where coherent structures are the most intense or the most frequent (Dupont and Brunet 2009). Figure \ref{qcriterion} show a time-averaged isosurface of positive Q-criterion over the canopy. 

\noindent Vortex tubes are visualized all along the top of the canopy, with a distinct pattern of Kelvin-Helmholtz waves near he leading edge, the transformation into vortex tubes aligned with the span-wise direction, and then the reorientation and break up of these vortical structures to create further instabilities, all consistent with findings from Quinn et al. (1995), Finnigan and Brunet (1995), and Dupont and Brunet (2009). Distinct structures are also apparent adjacent to the lateral and trailing edges of the canopy.

\subsubsection*{Vorticity Budget}

Vorticity budgets for several volumes throughout the domain are presented in this section. Regions where there are sharp gradients in surface area-to-volume ratio, $a_v$, in the canopy are examined to quantify the contributions and effects of the drag tilting and stretching term on the overall budget of vorticity in these locations. The region above the top surface of the canopy is also analyzed, where the turbulent coherent structures of similar length scales can be seen in Figure \ref{qcriterion}. Vorticity budgets are presented as follows: the fractional contribution of each direction is presented to understand which direction $(\xi, \eta, \zeta)$ dominates the vorticity in the region, then each term is presented as a fractional contribution to the absolute value total in the relevant direction. This allows the reader to visualize which terms are dominating as the simulation develops in time.

\begin{figure}[h!]
        \centering
        \begin{annotatedFigure}
        {\includegraphics[width=4.5cm]{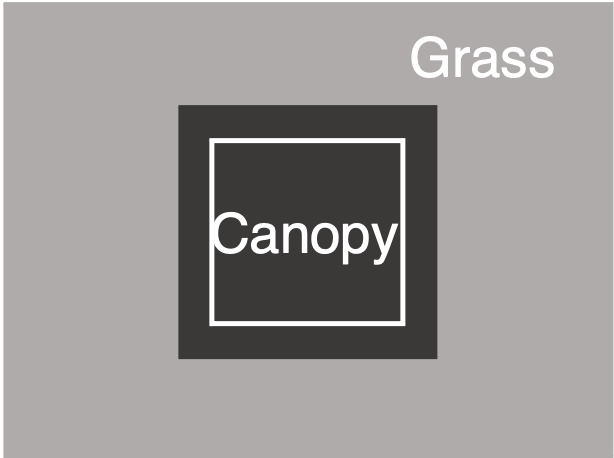}} 
        \annotatedFigureText{0.4985,0.99}{black}{0.0523}{\large A} 
        \end{annotatedFigure} \\
        \begin{annotatedFigure}
        {\includegraphics[width=8.15cm]{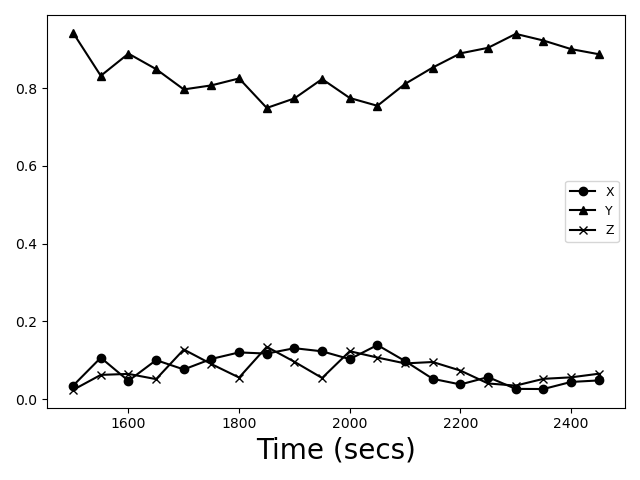}}
        \annotatedFigureText{0.4985,0.99}{black}{0.0523}{\large B}
        \end{annotatedFigure}
        \begin{annotatedFigure}
        {\includegraphics[width=8.15cm]{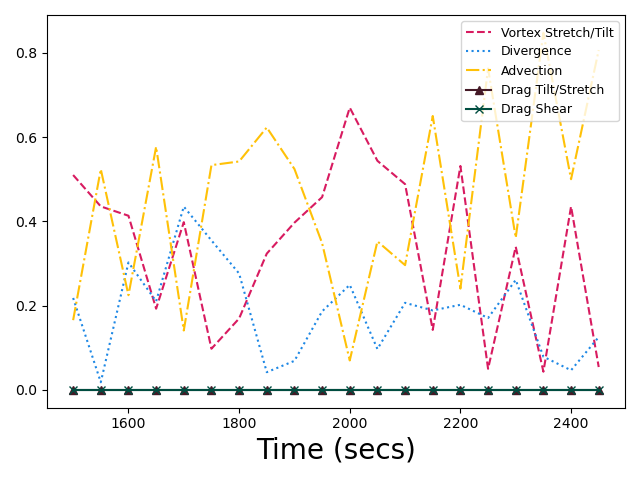}}
        \annotatedFigureText{0.4985,0.99}{black}{0.0523}{\large C}
        \end{annotatedFigure}
        \begin{annotatedFigure}
        {\includegraphics[width=8.15cm]{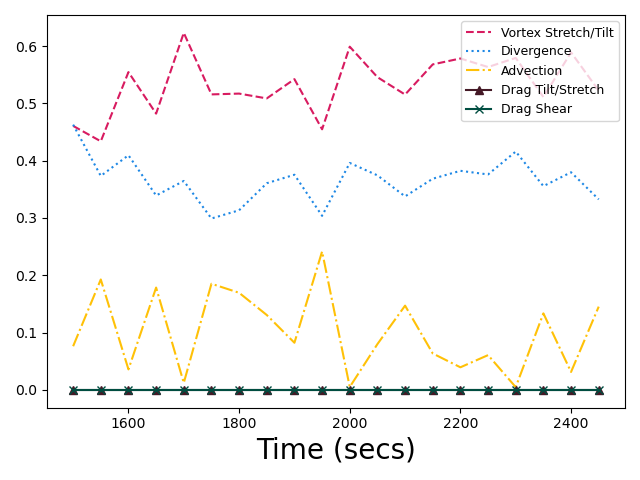}}
        \annotatedFigureText{0.4985,0.99}{black}{0.0523}{\large D}
        \end{annotatedFigure}
        \begin{annotatedFigure}
        {\includegraphics[width=8.15cm]{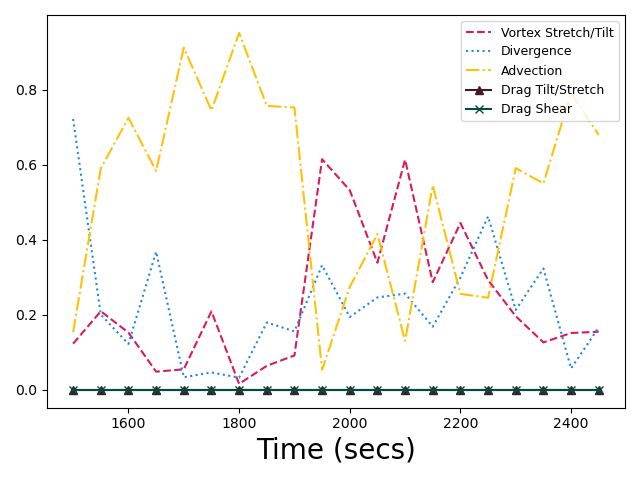}}
        \annotatedFigureText{0.4985,0.99}{black}{0.0523}{\large E}
        \end{annotatedFigure}
        \caption{Vorticity budgets in a volume above the top surface of the canopy by component. A top-down view of this volume is visualized in (A). The absolute value of the three normalized components of vorticity (B), $\xi$ (C), $\eta$ (D), and $\zeta$ (E), by term in the volume containing the the coherent structures of figure \ref{qcriterion} are presented.}
        \label{VB_topsurf}
\end{figure}

In Figure \ref{VB_topsurf}, a volume containing the coherent structures, above the canopy, observed in Figure \ref{qcriterion}, is analyzed. Figure \ref{VB_topsurf}(A) shows the top-down view of the volume above the canopy over which the terms were averaged. Figure \ref{VB_topsurf}(B) shows the volume averaged relative magnitudes of each component of vorticity normalized by total magnitude. Figures \ref{VB_topsurf}(C), \ref{VB_topsurf}(D), and \ref{VB_topsurf}(E) show the absolute value of individual terms of the component vorticity equations in $x$, $y$, and $z$, respectively. In all three components, the three dominant contributors of vorticity are the advection $\omega_{adv}$, divergence $\omega_{div}$, and the vortex tilting and stretching $\omega_{vts}$. Figure \ref{VB_topsurf}(B), shows that $\eta$ accounts for $> 70\%$ of the vorticity for all times in this window. Thus, we will focus our discussion on $\eta$ except to comment on the time rate of change, $\dot{\eta}$. It is observed that $\dot{\eta}$ is negatively correlated with $\dot{\zeta}$, suggesting that vorticity is exchanged between $\eta$ and $\zeta$. We see that when $\dot{\eta} < 0$, $\dot{\zeta} > 0$, and conversely when $\dot{\zeta} < 0$, $\dot{\eta} > 0$. As will be shown below, the exchange between components or terms is a common occurrence in this study. For the volume analyzed in Figure \ref{VB_topsurf}, this exchange happens through a balance of $\omega_{vts}$ and $\omega_{adv}$. In both \ref{VB_topsurf}(D) and \ref{VB_topsurf}(E), $\dot{\omega}_{vts}$ and $\dot{\omega}_{adv}$ are predominantly negatively correlated. Vortex stretching and tilting is the dominating term in $\eta$, as can be seen in Figure \ref{qcriterion}, as the vortices are stretching, breaking up, and tilting into the vertical direction in the shear instability layer. Divergence also plays a role in this volume from momentum being ejected upwards out of the porous media that represents the canopy. The drag terms, $\omega_{dts}$ and $\omega_{dshear}$ are zero because the volume is above the canopy and does not contain the top of the canopy.



\begin{figure}[h!]
        \centering
        \begin{annotatedFigure}
        {\includegraphics[width=4.75cm]{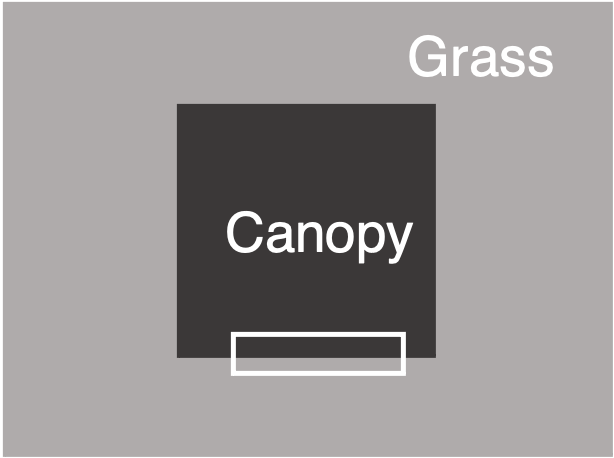}} 
        \annotatedFigureText{0.4985,0.99}{black}{0.0523}{\large A} 
        \end{annotatedFigure} \\
        \begin{annotatedFigure}
        {\includegraphics[width=8.15cm]{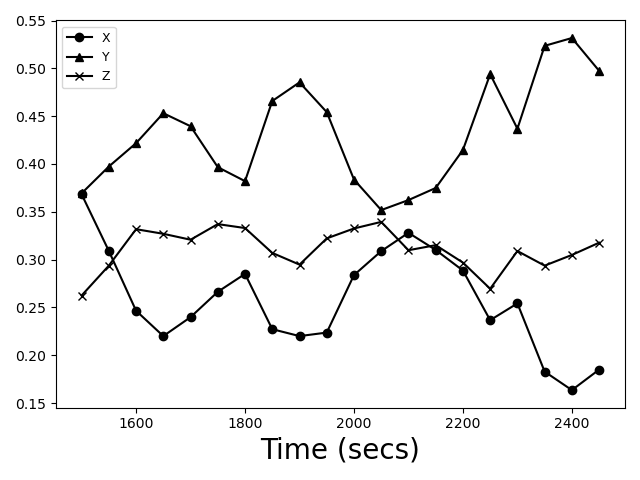}}
        \annotatedFigureText{0.4985,0.99}{black}{0.0523}{\large B}
        \end{annotatedFigure}
        \begin{annotatedFigure}
        {\includegraphics[width=8.15cm]{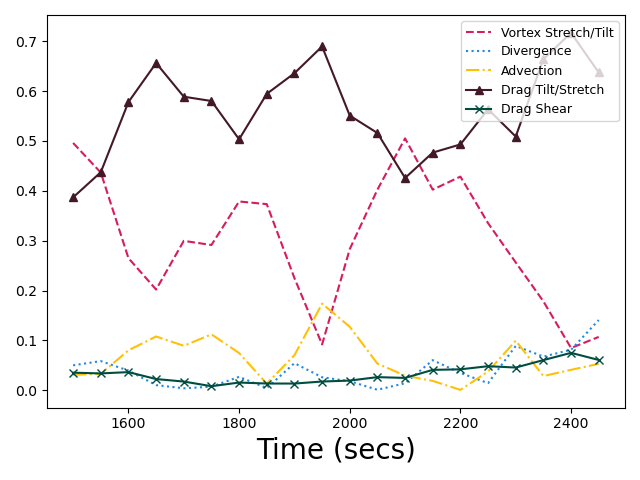}}
        \annotatedFigureText{0.4985,0.99}{black}{0.0523}{\large C}
        \end{annotatedFigure}
        \begin{annotatedFigure}
        {\includegraphics[width=8.15cm]{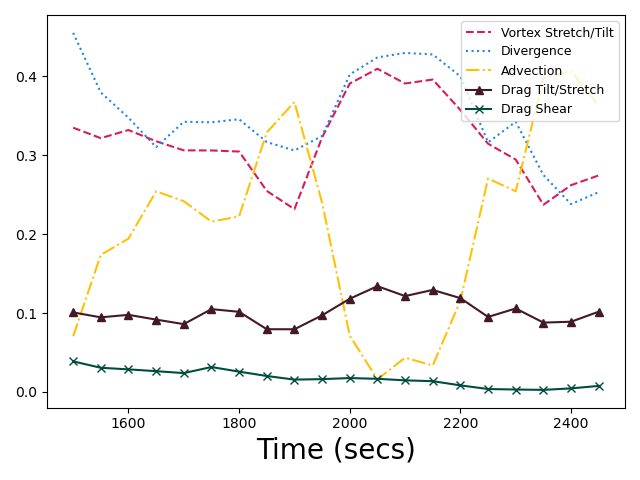}}
        \annotatedFigureText{0.4985,0.99}{black}{0.0523}{\large D}
        \end{annotatedFigure}
        \begin{annotatedFigure}
        {\includegraphics[width=8.15cm]{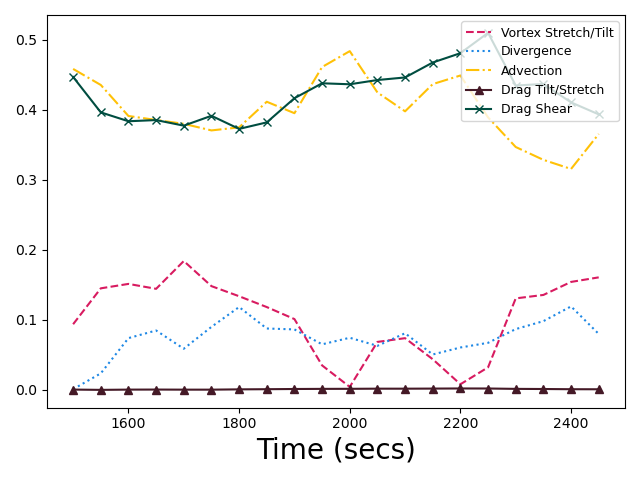}}
        \annotatedFigureText{0.4985,0.99}{black}{0.0523}{\large E}
        \end{annotatedFigure}
        \caption{Vorticity budgets at the bottom lateral edge by component. The volume containing the bottom lateral edge is visualized in (A). The absolute value of the three normalized components of vorticity (B), $\xi$ (C), $\eta$ (D), and $\zeta$ (E), by term in the volume containing the bottom lateral edge are presented.}
        \label{VB_latedge}
\end{figure}

The volume over the bottom lateral edge contains a sharp positive gradient in $a_v$ in the y-direction. Figure \ref{VB_latedge}A shows the volume at the bottom lateral edge of the canopy over which the terms were averaged. Figure \ref{VB_latedge} shows fractional contributions $(\xi,\eta,\zeta)$ to the total magnitude in \ref{VB_latedge}(B) and contributions of each term to the components: $\xi$ in \ref{VB_latedge}(C), $\eta$ in \ref{VB_latedge}(D), and $\zeta$ in \ref{VB_latedge}(E). Figure \ref{VB_latedge}(B) shows that $\eta$ is the largest component, but unlike the top surface, all three components have comparable magnitudes. We see that $\eta$ and $\xi$ are exchanging vorticity, indicated by their negatively correlated rates of change and $\zeta$ is oscillating about $30\%$ of the total. Figure \ref{VB_latedge}(C) shows that $\xi$ is heavily dominated by the drag tilting and stretching term, $\xi_{dts}$, followed by the vortex tilting and stretching, $\xi_{vts}$. It is also apparent that these terms are exchanging vorticity. As $\xi_{dts}$ increases in time, $\xi_{vts}$ decreases in time and vice versa.  This tilting and stretching accounts for the majority of the vorticity in $\xi$. However, the largest component of the vorticity in this volume is $\eta$. Figure \ref{VB_latedge}(D) shows a three-way balance of contributions from the $\eta_{vts}$, $\eta_{div}$, and $\eta_{adv}$ terms. The $\eta_{vts}$ and $\eta_{div}$ terms have very similar profiles and magnitudes as they evolve in time. The rates of change of $\eta_{adv}$ and $\eta_{div}$ are negatively correlated, suggesting an exchange of vorticity between these terms. In this volume, $\eta$ is dependent on the $u$ and $w$ velocities, as well as gradients in the $x$ and $z$ direction. The wind flow has a shear profile that grows in thickness downwind of the leading edge of the canopy. Thus, advection plays an important role because air is being transported along the x-direction. Vortex stretching and tilting, $\eta_{vts}$ is significant in this region because,as the shear boundary layer develops downstream of the leading edge of the canopy, vortices are being tilted and stretched creating greater instability further downwind. The z-component, $\zeta$, Figure \ref{VB_latedge}(E), at the lateral edge is dominated by the drag shear term, $\zeta_{dshear}$, as this component is heavily dominated by the shear boundary layer developing on the lateral edge of the canopy. As seen in Figure \ref{streamline_vort}, the volume containing the lateral edge displays vortex lines hugging the lateral surface of the canopy where $\zeta_{dshear}$ is a large contributor to $\zeta$. The vorticity budget for the top lateral edge was also examined and was found to behave analogous to the bottom lateral edge. Hence, those results are not presented.


\begin{figure}[h!]
        \centering
        \begin{annotatedFigure}
        {\includegraphics[width=4.75cm]{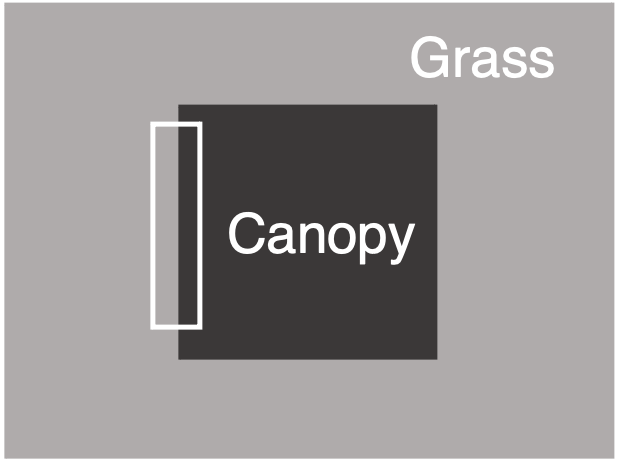}} 
        \annotatedFigureText{0.4985,0.99}{black}{0.0523}{\large A} 
        \end{annotatedFigure} \\
        \begin{annotatedFigure}
        {\includegraphics[width=8.15cm]{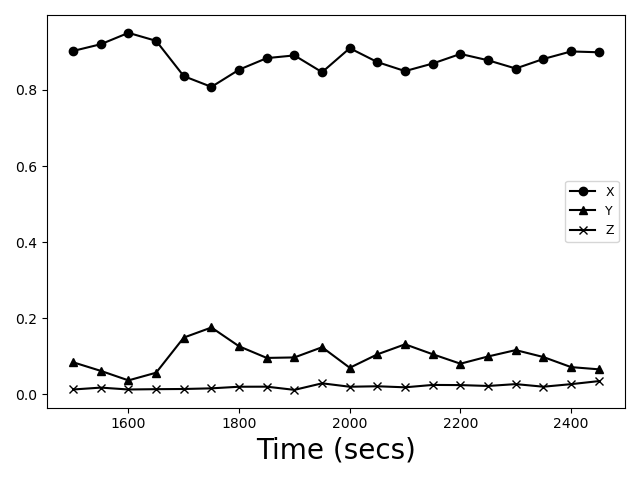}}
        \annotatedFigureText{0.4985,0.99}{black}{0.0523}{\large B}
        \end{annotatedFigure}
        \begin{annotatedFigure}
        {\includegraphics[width=8.15cm]{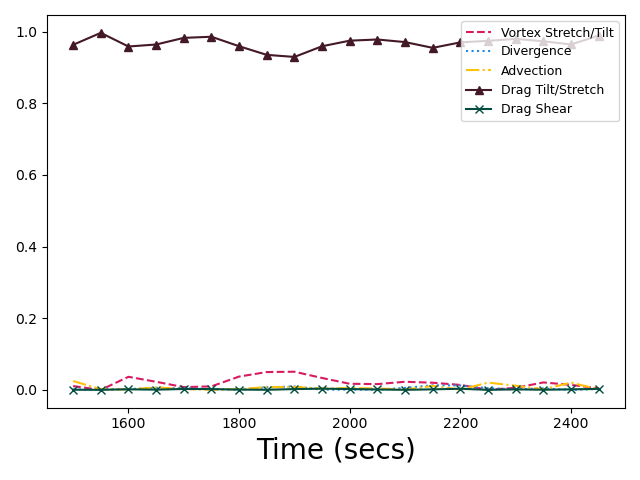}}
        \annotatedFigureText{0.4985,0.99}{black}{0.0523}{\large C}
        \end{annotatedFigure}
        \begin{annotatedFigure}
        {\includegraphics[width=8.15cm]{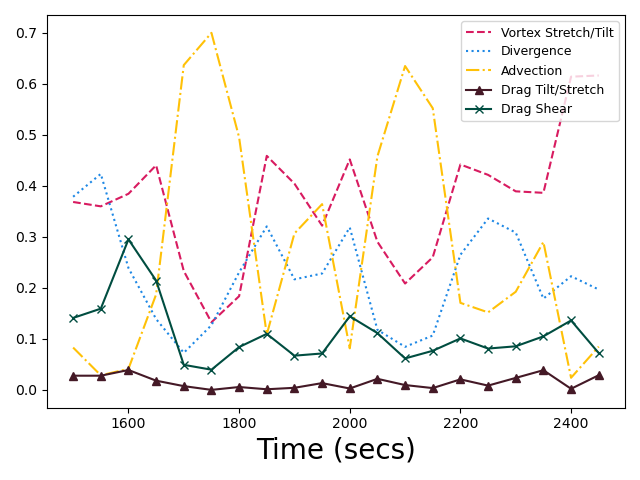}}
        \annotatedFigureText{0.4985,0.99}{black}{0.0523}{\large D}
        \end{annotatedFigure}
        \begin{annotatedFigure}
        {\includegraphics[width=8.15cm]{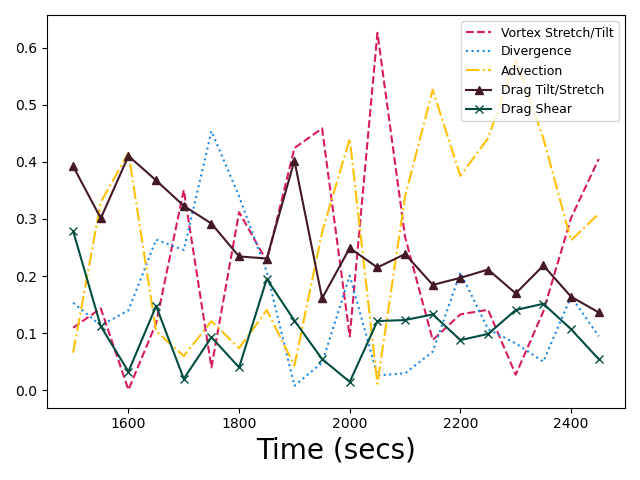}}
        \annotatedFigureText{0.4985,0.99}{black}{0.0523}{\large E}
        \end{annotatedFigure}
        \caption{Vorticity budgets at the leading edge by component. The volume containing the leading edge is visualized in (A).The absolute value of the three normalized components of vorticity (B), $\xi$ (C), $\eta$ (D), and $\zeta$ (E), by term in the volume containing the leading edge are presented.}
        \label{VB_leadedge}
\end{figure}
The vorticity budget over the volume containing the leading edge of the canopy is presented in Figure \ref{VB_leadedge} where the same quantities displayed in Figures \ref{VB_topsurf} and \ref{VB_latedge} are plotted. Figure \ref{VB_leadedge}(A) shows the volume at the leading edge of the canopy over which the terms were averaged.
Figure \ref{VB_leadedge}(B) shows that $\xi$ completely dominates the vorticity field, representing $ > 80\%$ of the total the majority of the time. In the x-direction, $xi$, Figure \ref{VB_leadedge}(C), the vorticity budget is completely dominated by the drag tilting and stretching term $\xi_{dts}$. This is expected because of the gradient from the grass to the canopy, as well as the positive vertical velocity that occurs in this region as the flow goes up over the canopy. In the y-direction, $\eta$, Figure \ref{VB_leadedge}(D), the budget is predominantly a balance of the divergence term, $\eta_{div}$, the vortex tilting and stretching term, $\eta_{vts}$, and the advection term, $\eta_{adv}$. The divergence of the flow becomes important because of the streamwise direction of the flow combined with the presence of the canopy moving air around the vegetation in the canopy. The vorticity budget in the z-direction, $\zeta$, Figure \ref{VB_leadedge}(E), shows the trade-off between the majority of terms, where the drag tilting and stretching, $\zeta_{dts}$, vortex tilting and stretching, $\zeta_{vts}$, and the advection, $\zeta_{adv}$ dominating. These three terms play an important role in $\zeta$ at the leading edge: the drag tilting and stretching because of the gradient in $a_v$, the vortex stretching and tilting because of the tilting and stretching of the vortex sheet in the shear boundary layer, which is also dominating in the y-component, and advection because the flow is being redirected vertically with the vertical positive velocity right before the leading edge.


\begin{figure}[h!]
        \centering
        \begin{annotatedFigure}
        {\includegraphics[width=4.75cm]{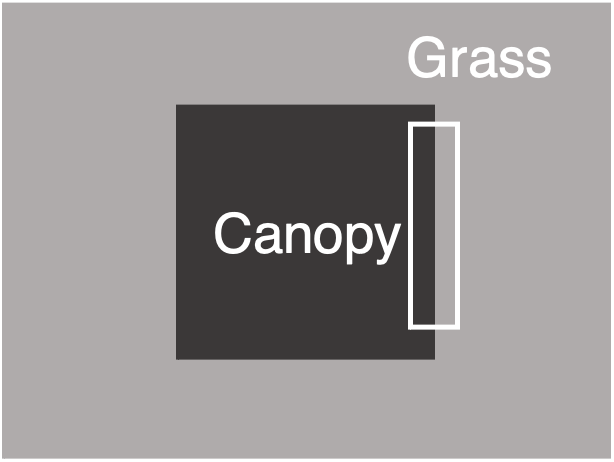}} 
        \annotatedFigureText{0.4985,0.99}{black}{0.0523}{\large A} 
        \end{annotatedFigure} \\
        \begin{annotatedFigure}
        {\includegraphics[width=8.15cm]{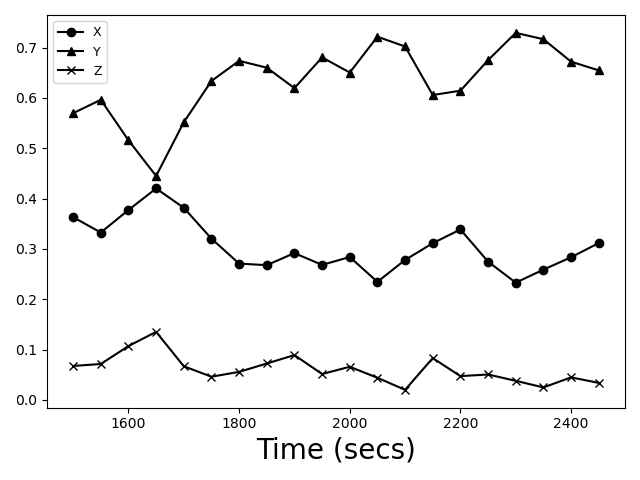}}
        \annotatedFigureText{0.4985,0.99}{black}{0.0523}{\large B}
        \end{annotatedFigure}
        \begin{annotatedFigure}
        {\includegraphics[width=8.15cm]{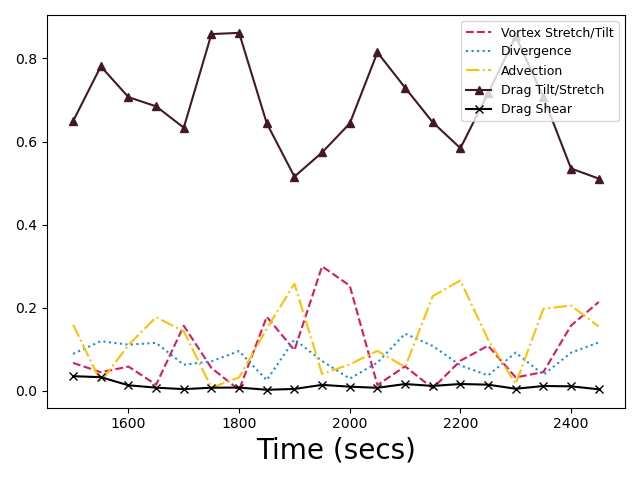}}
        \annotatedFigureText{0.4985,0.99}{black}{0.0523}{\large C}
        \end{annotatedFigure}
        \begin{annotatedFigure}
        {\includegraphics[width=8.15cm]{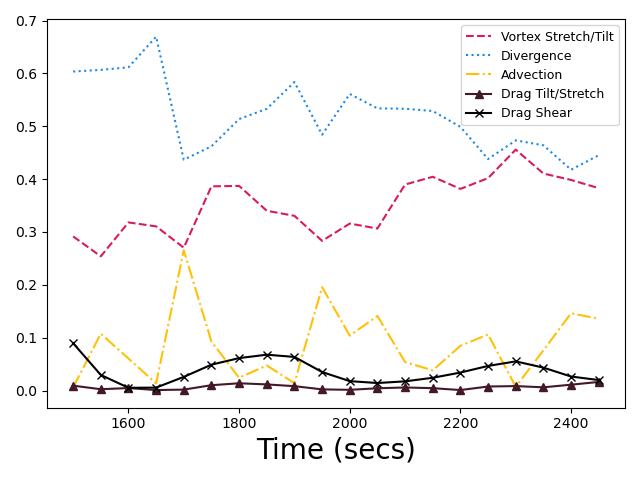}}
        \annotatedFigureText{0.4985,0.99}{black}{0.0523}{\large D}
        \end{annotatedFigure}
        \begin{annotatedFigure}
        {\includegraphics[width=8.15cm]{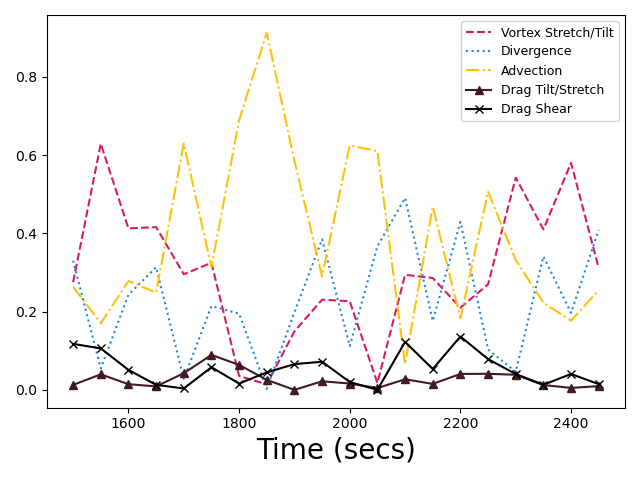}}
        \annotatedFigureText{0.4985,0.99}{black}{0.0523}{\large E}
        \end{annotatedFigure}
        \caption{Vorticity budgets at the trailing edge by component. The volume containing the trailing edge is visualized in (A).The absolute value of the three normalized components of vorticity (B), $\xi$ (C), $\eta$ (D), and $\zeta$ (E), by term in the volume containing the trailing edge are presented.}
        \label{VB_trail}
\end{figure}

Finally, the budget over a volume containing the trailing edge is presented in Figure \ref{VB_trail}. Figure \ref{VB_trail}(A) shows the volume at the trailing edge of the canopy over which the terms were averaged. Once again, the same quantities are plotted in Figure \ref{VB_trail} as in the previous three figures. The vorticity budgets for $\xi$ and $\eta$ over this volume are similar to the leading edge in the dominant terms. The difference lies in the z-component, where $\zeta_{dts}$ plays a very small role at the trailing edge. The presence of the leading edge is what causes the vertical orientation of vorticity up and over the canopy. The trailing edge is surrounded by well-developed turbulence and, thus, the sharp gradient in $a_v$ does not contribute as much as the leading edge to the vorticity budget in this region. Throughout the different regions surrounding the canopy in which vorticity budgets were analyzed, certain terms played a lesser role. The baroclinicity remained near zero as is expected since we are not considering wildfires or heating at the surface that would induce horizontal temperature/density gradients. In other words, the absence of heat means there is no mixing of warm and cold air. So, the density and pressure gradients do not significantly separate. The turbulent viscosity is also very small compared to the other terms, as the residual stress tensor-eddy viscosity relation used in the Smagorinsky model represents the subgrid-scale turbulence, which accounts for a small percentage of the total turbulence and vorticity in the system. Therefore, $\omega_{bar}$ and $\omega_{visc}$ were omitted from the plots.



\section*{Conclusion}
The development of large coherent structures in canopy flow has been discussed and described in the literature for decades. There is a widely-accepted understanding of the generation of a boundary layer above the canopy that forms from the transformation of the initial Kelvin-Helmholtz instability caused by wind shear at the leading edge of the canopy. These coherent structures are present in areas of high vorticity, yet quantifying the vorticity in these areas has yet to be emphasized in the literature.

In an effort to quantify the vorticity that is qualitatively described across many studies, we derived a vorticity transport equation in HIGRAD/FIRETEC. In order to derive the vorticity budget equation, the Smagorinsky eddy viscosity parameterization was used to model the sub-filter scale turbulence and the results of this implementation is  described in part 1 of this study, (\textit{Perez et al. (2024), submitted}). In the present study, we described the role of vorticity in canopy wind flow and quantify the vorticity near the canopy. It is important to note that the vorticity and momentum equations conserve the same thing: momentum. Vorticity is a surrogate for angular momentum. The utility of analyzing vorticity and its governing equation is that it gives a different perspective on the flow field. The curl of the conservation of momentum equation yields the vorticity equation, a tensor gradient of velocity. This equation permits one to observe the sources and sinks to the velocity gradients.

In the momentum equation, the drag term is purely a sink to momentum. The curl of the drag term appears in the vorticity equation as two terms. The derivation of the vorticity budget equation yielded a two-part drag term, whose first part is referred to here as the drag tilting and stretching term, $\omega_{dts}$, and the second part is referred to as the drag shear term, $\omega_{dshear}$. The derivation of these terms highlighted the way the vorticity of the system changes due to the geometry and characteristics of the canopy, specifically gradients of the surface area-to-volume ratio, $a_v$.  Surprisingly, the drag and stretching term tilting term, is a \textit{source} term. That is, the momentum drag contains a velocity gradient source and hence a vorticity source. Specifically, it is the gradient in the effective drag that sources vorticity. The second term is purely a vorticity sink. 

Examining the high-vorticity generating areas around the canopy revealed coherent structures present in areas previously stated in the literature: at the leading edge of the canopy, in the shear boundary layer, and downwind of the canopy. Areas of high vorticity magnitude near the sharp gradients in $a_v$, namely at the edges of the canopy, were also discovered and analyzed. Vorticity-relevant statistics and a vorticity budget were presented for these regions. Taking a closer look at the drag tilting and stretching term, $\omega_{dts}$, confirms the way vorticity is transferred from the horizontal to the vertical components of the vorticity at the lateral edges and top, causing the tilting of the vortical rotation axis. We also described how the same term concentrates vorticity on vegetation interfaces due to stretching. Thus, the lateral and top edges are an area of high vorticity magnitude. The vorticity budget over this area shows the importance of the drag tilting and stretching term in the total transport of vorticity compared to other terms for canopy flow. The drag tilting term is not the largest term, but it is a crucial source to vorticity, which then evolves through the nonlinearities in the other terms.

In wildfire dynamics, ambient vorticity is often obscured by other vortical phenomena like fire whirls and horizontal roll vortices. Understanding the ambient vorticity around a canopy plays an important role in being able to anticipate the conditions that enhance these other fire-based phenomena because fire not only interacts with and modifies the ambient vorticity, it also generates additional vorticity (Forthofer and Goodrick 2011). The introduction of heat and aerosols from a fire introduce horizontal gradients in temperature and density, complicating the analysis of the vorticity budget. 

This study is the first of its kind in quantifying the ambient vorticity through the use of a vorticity budget equation in canopy flow. Potter (2012) highlighted the importance in understanding where vorticity equation terms dominate as well as understanding where ambient vorticity takes over compared to vorticity created by buoyancy gradients (Potter 2012). Vorticity in wildfire has been investigated and presented in the literature in depth. Specifically horizontal vorticity generation due to buoyancy gradients and the reorientation of this vorticity by entrainment (Potter 2012). Vorticity has been extensively described in terms of plume dynamics and fire behavior (Potter 2012). The ambient vorticity characterized and quantified in the present investigation has direct implications on the dynamics introduced by a heat source or wildfire, and is beyond the scope of this study. This study serves as a first step in understanding and gaining some insight in the ambient vorticity dynamics that can have implications on wildfire behavior.

We hypothesize that the introduction of a heat source would cause the baroclinicity term in the vorticity budget to play a greater role, but this has yet to be studied (Forthofer and Goodrick 2011). Under the right conditions, this could lead to the development of fire whirls, a feature  yet to be investigated using HIGRAD/FIRETEC. The addition of heat would alter the vorticity lines, causing them to curve in the vertical direction, serving as axes of rotations for vortices. This process is further described in Forthofer and Goodrick (2011). We conclude that although this study presents valuable initial findings, there are still many gaps in the knowledge of ambient vorticity in canopy flow, as well as the direct implications this has on wildfire dynamics in practice. \\
\\
\noindent \textbf{Dorianis M. Perez:} Conceptualization, Methodology, Software, Formal Analysis, Writing - Original Draft, Visualization, Writing - Review \& Editing; \textbf{Jesse Canfield:} Conceptualization, Writing - Review \& Editing, Supervision; \textbf{Rodman Linn:} Writing - Review \& Editing, Funding Acquisition; \textbf{Kevin Speer:} Conceptualization, Writing - Review \& Editing 

\section*{Acknowledgements}
This work was supported by the U.S. Department of Energy through the Los Alamos National Laboratory. Los Alamos National Laboratory is operated by Triad National Security, LLC, for the National Nuclear Security Administration of U.S. Department of Energy (Contract No. 89233218CNA000001). We express our sincere gratitude to the U.S. Department of Defense, Strategic Environmental Research and Development Program (SERDP) for partially funding this study under grant RC20-C3-1298. We additionally thank Los Alamos National Laboratories Center for Space and Earth Science (CSES) for partially funding this work under LDRD project 20240477CR-SES. All computations for this work were performed using Los Alamos National Laboratory's Institutional Computing Resources. 

\section*{References}

\hspace{0.2in} 
Bebieva, Y., Speer, K., White, L., Smith, R., Mayans, G., Quaife, B. (2021) Wind in a natural and artificial wildland fire fuel bed. Fire 4, 30. \\

Benahmed, L., Aliane, K. (2019) Simulation and analysis of a turbulent flow around a three-dimensional obstacle. acta mechanica et automatica 13(3), 173-180. \\

Brunet, Y., Finnigan, J.J., Raupach, M.R. (1994) A wind tunnel study of air flow in waving wheat: single-point velocity statistics. Boundary Layer Meteorology 70, 95–132. \\

Canfield, J.M., Linn, R.R., Cunningham, P., Goodrick, S.L. (2005) Modelling effects of atmospheric stability on wildfire behaviour. In ‘Proceedings 6th Fire and Forest Meteorology Symposium and 19th Interior West Fire Council Meeting’, 25–27 October 2005, Canmore, AB, Canada. \\

Canfield, J.M., Linn, R.R., Sauer, J.A., Finney, M., Forthofer, J. (2014) A numerical investigation of the interplay between fireline length, geometry, and rate of spread. Agricultural and Forest Meteorology 189-190, 48-59. \\

Colman, J. J., Linn, R. R. (2007) Separating combustion from pyrolysis in HIGRAD/FIRETEC. International Journal of Wildland Fire 16(4), 493–502. \\

Cunningham P., Goodrick S. L., Hussaini M. Y., Linn R. R. (2005) Coherent vortical structures in numerical simulations of buoyant plumes from wildland fires. International Journal of Wildland Fire 14, 61-75. \\

Denaro, F.M. (2018) What is the basic physical meaning of enstrophy? URL \\ 
https://www.researchgate.net/post/What\_is\_the\_basic\_physical\_meaning\_of\_enstrophy. \\

Dupont, S., Brunet, Y. (2007) Edge flow and canopy structure: a large-eddy simulation study. Boundary-Layer Meteorology 126, 51–71. \\

Dupont, S., Brunet, Y. (2008b) Influence of foliar density profile on canopy flow: a large-eddy simulation study. Agricultural and Forest Meteorology 148, 976–990. \\

Dupont, S., Brunet, Y. (2009) Coherent structures in canopy edge flow: a large-eddy simulation study. Journal of Fluid Mechanics 630, 93-128. \\

Dwyer, M.J., Patton, E.G., Shaw, R.H. (1997) Turbulent kinetic energy budgets from a large eddy simulation of airflow above and within a forest. Boundary Layer Meteorology 84, 23–43 \\

Finnigan, J.J. (1979) Turbulence in waving wheat. Boundary Layer Meteorology 161, 181-211. \\

Finnigan, J.J. (1979) Turbulence in waving wheat. II. Structure of momentum transfer. Boundary Layer Meteorology 161, 213-236. \\

Finnigan, J. J. (2000) Turbulence in plant canopies. Annual Revision of Fluid Mechanics 32, 519–571. \\

Finnigan, J. J., Brunet, Y. (1995) Turbulent airflow in forests on flat and hilly terrain. In Wind and Trees (ed. M. P. Coutts \& J. Grace), pp. 3–40. Cambridge University Press. \\

Forthofer, J.M., Goodrick, S.L. (2011) Review of Vortices in Wildland Fire. Journal of Combustion, 984363, 1-14. \\

Gao, W., Shaw, R.H., Paw, U.K.T. (1989) Observation of organised structures in turbulent flow within and above a forest canopy. Boundary Layer Meteorology 47(1), 349–377. \\

Hall, W.D. (1980) A detailed microphysical model within a two-dimensional dynamic framework: model description and preliminary results. Journal of Atmospheric Science, 37, 2486-2507. \\

Holton, J.R., Hakim, G.J. (2012) An Introduction to Dynamic Meteorology, 5th ed. Amsterdam: Academic Press. \\

Ishiyama, R., Tanaka, H.L. (2021) Analysis of Vorticity Budget for a Developing Extraordinary Arctic Cyclone in August 2016. SOLA 17: 120-124. \\

Jeong, J, Hussain, F. (1989) On the identification of a vortex. Journal of Fluid Mechanics
285, pp. 69–94.  \\

Josephson, A. J., Casta$\tilde{n}$o, D., Koo, E., Linn, R. R. (2020) Zonal-Based Emission Source Term Model for Predicting Particulate Emission Factors in Wildfire Simulations. Fire Technology \\

Kaimal, J.C., Finnigan, J.J. (1994) Atmospheric Boundary Layer Flows: Their Structure and Measurement. New York: Oxford University Press. \\

Kanda, M., Hino, M. (1994) Organized structures in developing turbulent-flow within and above a plant canopy, using a LES. Boundary Layer Meteorology 68(3), 237–257. \\

Katul, G.G. (1997) The ejection-sweep character of scalar fluxes in the unstable surface layer. Boundary Layer Meteorology, 83, 1-26. \\

Koo, E., Linn, R.R., Pagni, P.J., Edminster, C.B. (2012) Modelling firebrand transport in wildfires using HIGRAD/FIRETEC. International Journal of Wildland Fire 21(4), 396-417. \\

Leclerc, M.Y., Beissner, K.C., Shaw, R.H., Den Hartog, G., Neumann, H.H. (1990) The influence of atmospheric stability on the budgets of the Reynolds stress and turbulent kinetic energy within and above a deciduous forest. Journal of Applied Meteorology 29, 916–933. \\

Lesnik, G. E. (1974) Results of measurement of turbulent energy balance components in a layer of vegetation. Izv. Atmos. Oceanic Phys. 10, 652–655. \\

Linn, R.R. (1997) A transport model for prediction of wildfire behavior. Los Alamos National Laboratory, Science Report LA-13334-T. (Los Alamos, NM) \\

Linn, R. R., Anderson, K., Winterkamp, J. L., Brooks, A., Wotton, M., Dupuy, J.-L., . . . Edminster, C. (2012) Incorporating field wind data into FIRETEC simulations of the International Crown Fire Modeling Experiment (ICFME): preliminary lessons learned. Canadian Journal of Forest Research 42 (5), 879–898. \\

Linn, R. R., Winterkamp, J. L., Colman, J. J., Edminster, C., Bailey, J. D. (2005) Modeling interactions between fire and atmosphere in discrete element fuel beds. International Journal of Wildland Fire 14(1), 37–48. \\

Lu, C.H., Fitzjarrald, D.R. (1994) Seasonal and diurnal variations of coherent structures over a deciduous forest. Boundary-Layer Meteorology 69(1–2), 43–69. \\

Meyers, T.P., Baldocchi, D.D. (1991) The budgets of turbulent kinetic energy and Reynolds stress within and above a deciduous forest. Agricultural and Forest Meteorology 53, 207–222. \\

Patton, E.G., Shaw, R.H., Judd, M.J., Raupach, M.R. (1998) Large-eddy simulation of windbreak flow. Boundary Layer Meteorology 87, 275–307. \\

Pimont, F., Dupuy, J.L., Linn, R.R., Dupont, S. (2009) Validation of FIRETEC wind-flows over a canopy and a fuel-break. International Journal of Wildland Fire 18(7), 775. \\

Pimont, F., Dupuy, J.L., Linn, R.R., Dupont, S. (2011) Impacts of tree canopy structure on wind flows and fire propagation simulated with FIRETEC. Annals of Forest Science 68(3), 523–530. \\

Pimont. F., Dupuy, J.L., Linn, R.R. "Wind and Canopies" In \textit{Wildland Fire Dynamics: Fire Effects and Behavior from a Fluid Dynamics Perspective}, ed. Speer, K., Goodrick, S. (Cambridge University Press: 2022), pages 183-208. \\
 
Pope, S. B. (2000) Turbulent Flows. Cambridge University Press. \\ 
 
Potter, B.E. (2012) Atmospheric interactions with wildland fire behaviour – II. Plume and vortex dynamics. International Journal of Wildland Fire 21, 802-817. \\

Quine, C. P., Coutts, M. P., Gardiner B., Pyatt, D. G. (1995) Forests and wind: management to minimize damage. In Forestry Commission Bulletin 114. (ed. B. Gardiner), HMSO Publications Centre. \\

Raupach, M., Copping, P., Legg, B. (1986) Experiments on scalar dispersion within a model plant canopy part I: The turbulence structure. Boundary Layer Meteorology 35, 21-52. \\

Raupach, M.R., Finnigan, J.J., Brunei, Y. (1996) Coherent eddies and turbulence in vegetation canopies: The mixing-layer analogy. Boundary Layer Meteorology 78, 351–382. \\

Raupach, M.R., Shaw, R.H. (1982) Averaging procedures for flow within vegetation canopies. Boundary Layer Meteorology 22, 79–90.

Reisner JM, Wynne S, Margolin L, Linn RR (2000a) Coupled atmospheric– fire modeling employing the method of averages. Monthly Weather Review 128, 3683–3691.  \\

Reisner JM, Knoll DA, Mousseau VA, Linn RR (2000b) New numerical approaches for coupled atmosphere–fire models. In ‘Proceedings of the Third Symposium on Fire and Forest Meteorology’, January 2000, Long Beach, CA. pp. 11–14. (American Meteorology Society: Boston, MA). \\

Saeedipour, M. (2023) An enstrophy-based analysis of the turbulence-interface interactions across the scales. International Journal of Multiphase Flow 164: 104449. \\

Schröder, A., Willert, C., Schanz, D. Geisler, R., Jahn, T., Gallas, Q., Leclaire, B. (2020) The flow around a surface mounted cube: a characterization by time-resolved PIV, 3D Shake-The-Box and LBM simulation. Exp Fluids 61, 189. \\

Sharples, J.J., Hilton, J.E., Badlan, R.L., Thomas, C.M., McRae, R.H.D. "Fire Line Geometry and Pyroconvective Dynamics" In \textit{Wildland Fire Dynamics: Fire Effects and Behavior from a Fluid Dynamics Perspective}, ed. Speer, K., Goodrick, S. (Cambridge University Press: 2022), pages 77-128. \\

Shaw R.H., Schumann, U. (1992) Large-eddy simulation of turbulent flow above and within a forest. Boundary Layer Meteorology 61, 47–64. \\

Shaw, R.H., Seginer, I. (1985) The dissipation of turbulence in plant canopies. Extended abstract: American Meteorological Society 7th Symposium on Turbulence and Diffusion Abstracts, pp 200–2033. \\

Shaw, R.H., Tavangar, J., Ward, D.P. (1983) Structure of the Reynolds stress in a canopy layer. Journal of Applied Meteorology and Climatology 22, 1922-1931. \\

Su, H.B., Shaw, R.H., Paw, U.K.T., Moeng, C.H., Sullivan, P.P. (1998) Turbulent statistics of neutrally stratified flow within and above a sparse for- est from large-eddy simulation and field observations. Boundary Layer Meteorology 88, 363–397. \\

Su, H.B., Shaw, R.H., Paw, U.K.T. (2000) Two-point correlation analysis of neutrally stratified flow within and above a forest from large-eddy simulation. Boundary Layer Meteorology 94, 423–460. \\

Tieszen, S.R. (2001) On the fluid mechanics of fires. Annual Review of Fluid Mechanics 33(1), 67–92. \\

Tohidi, A., Gollner, M.J.,  Xiao, H. (2018) Fire Whirls. Annual Review of Fluid Mechanics 50: 187-213. \\

Watanabe, T. (2004) Large-eddy simulation of coherent turbulence structures associated with scalar ramps over plant canopies. Boundary Layer Meteorology 112, 307–341. \\

Weiss, J. (1990) The dynamics of enstrophy transfer in two-dimensional hydrodynamics. Physica D 48: 273-294. \\

Wilson, N. R. and Shaw, R. H. (1977) A higher order closure model for canopy flow. Journal of Applied Meteorology 16, 1197–1205. \\

Yang, B., Raupach, M., Shaw, R.H., Paw, U.K.T., Morse, A.P. (2006a) Large-eddy simulation of turbulent flows across a forest edge. Part I: Flow statistics. Boundary Layer Meteorology 120, 377–412. \\

Yang, B., Morse, A.P., Shaw, R.H., Paw, U.K.T. (2006b) Large-eddy simulation of turbulent flow across a forest edge. Part II: Momentum and turbulence kinetic energy budgets. Boundary Layer Meteorology 121, 433–457. \\

Zhu, Y., Antonia, R.A. (1997) On the correlation between enstrophy and energy dissipation rate in a turbulent wake. Applied Scientific Research 57: 337-347.

\end{document}